\documentclass[a4paper,final,12pt,onecolumn]{article}
\usepackage{setspace}
\usepackage{listings}
\usepackage[cp1250]{inputenc}
\usepackage{graphicx}
\usepackage{mathrsfs}
\usepackage{url}
\usepackage{amssymb}
\usepackage{amsmath}
\usepackage{amsthm,amscd,amsfonts}

\newcommand{\vecw}{{\bf w}}
\newcommand{\vecs}{{\bf s}}

\title{Phoneme discrimination using $KS$-algebra II.}

\begin{document}

\author{Ondrej~\v Such\thanks{O.\v Such is with Slovak Academy of Sciences, Bansk\'a Bystrica, Slovakia, {\tt ondrejs@savbb.sk}}, Lenka Mackovičová\thanks{L. Mackovičová is with University of Matej Bel, Banská Bystrica, Slovakia, {\tt lenka.mackovicova@umb.sk}}%
\thanks{Work on this paper was partially supported by research grant VEGA 2/0112/11;
Computations were done on computers purchased in project ITMS code: 26210120002}%
}%
\maketitle

\begin{abstract}
$KS$-algebra consists of expressions constructed with four kinds operations, the minimum, maximum, difference and additively homogeneous generalized means. Five families of $Z$-classifiers are investigated on binary classification tasks between English phonemes. It is shown that the classifiers are able to reflect well known formant characteristics of vowels, while having very small Kolmogoroff's complexity.
\end{abstract}

\section{Introduction}

In our previous paper in the series we have  proposed a new $KS$--algebra for constructing binary phoneme classifiers based on spectral content. The algebra consists of expressions constructed from a vector of spectral values $\vecs = (s_1,\ldots, s_n)$, and the zero value by means of the following operators
\begin{itemize}
\item the minimum $\min(x_1, \ldots, x_n)$,
\item the maximum $\max(x_1, \ldots, x_n)$,
\item the difference $x_1 - x_2$,
\item the additively homogeneous means  $A_\alpha$, 
\end{itemize}
where 
\begin{align*}
A_\alpha(x_1, \ldots, x_n) &= 
\ln\Bigl(M_\alpha\bigl(\exp(x_1), \ldots, \exp(x_n)\bigr)\Bigr), 
\intertext{and $M_\alpha$ is the generalized mean}
M_\alpha(x_1, \ldots, x_n) &= \Bigl( \dfrac{x_1^\alpha + \cdots + x_n^\alpha}{n} \Bigr)^{1/\alpha}.
\end{align*}
In this article we shall present results of search for optimal $Z$-classifier in a large, albeit special family of elements of $KS$-algebra.

\section{Optimization setup}

For dataset we shall use spectral data presented in  \cite{EnglishPhonemes} and used for demonstration in \cite{esl} and \cite{HastieBuja}. The data is derived from TIMIT database, often used in speech recognition tasks. It consists of 5 English phonemes, three vowels {\tt aa, ao, iy} and two consonants {\tt dcl, sh}, each pronounced by a male speaker from various geographical regions. The sound was sampled at 16kHz, and spectral data was prepared using 512-sample window, resulting in 256 spectral vector for each sample. The data is divided into train and test categories, with approximately equal proportions in each. 

Let us recall that a general $Z$-classifier for phonemes $\phi_1, \phi_2$ corresponds to an element $f$ of $KS$-algebra. Suppose the classifier is presented with spectral data $\vecs$ and prior knowledge that the data corresponds to either $\phi_1$ or $\phi_2$. It decides that phoneme is $\phi_1$ if $f(\vecs) <0$ and decides that the phoneme is $\phi_2$ if $f(\vecs)>0$.

For an optimization criterion we chose the number of successful classifications $c(f)$ on the training data set. Since the data set is rather small, ties may occur. In the case of ties, we choose the classifier that 
maximizes the expression
$$
\rho(f) := \min\Bigl(\frac{\mu_1^2}{\sigma_1^2}, \frac{\mu_2^2}{ \sigma_2^2}\Bigr),
$$
where $\mu_i$, $\sigma_i$ are sample means and standard deviations for values of $f$ on the set of training samples of   phoneme $\phi_i$. The expression plays role analogous to that of Fischer's linear discriminant.

Since there is no obvious shortcut to finding an optimum, we resort to evaluating classification performance in turn for every classifier in a given family. 

\section{Families of classifiers}

By a \emph{spectral range} we mean a sequence $R_{i,j} = (s_i, s_{i+1}, \ldots, s_j)$ of consecutive spectral amplitudes (ordered by increasing frequency). Since brain structures devoted to speech recognition are tonotopically organized \cite{wiki:Tonotopy}, we propose to use for discrimination  functions defined on spectral ranges. Each discrimination function takes the form 
$$
f = f_1(R_{i,j}) - f_2(R_{k,l}),
$$
where $f_1,f_2$ are symmetric, additively homogeneous functions of $KS$-algebra. The difference of values of $f_1$ and $f_2$ is then intensity invariant. We distinguish five different classes of such functions
\begin{itemize}
\item[1.] the mean of values in the spectral range 
\item[2.] the mean of $m$ largest values in the spectral range
\item[3.] $A_1$ average of $m$ largest values in the spectral range
\item[4.] $A_2$ average of $m$ largest values in the spectral range
\item[5.] a quantile of the spectral range (the $m$-th largest value)
\end{itemize}
Obviously, family 1 is a subset of family 2. Families 2--5 can be seen as special cases of a family obtained by taking average $A_\alpha$ of $m$ largest values ($\alpha=0,1,2,-\infty$ respectively). Families 1, 2 and 5 are special cases of so called OWA-operators \cite{OWA}. A general $n$-ary OWA operator $F$ with weight vector $\vecw = (w_1, \ldots, w_n)$ is defined by expression
$$
F(a_1, \ldots, a_n) = w_1 b_1 + \cdots + w_nb_n,
$$
where $b_i$ is the $i$-th largest element of the set $\{a_1,\ldots, a_n\}$.

If one limits oneself to searching over pairs of functions defined on ranges of width up to $w$, the complexity of selecting the best one is $\sim kN w^6$ in families 2--5, and $\sim k' N w^4$ in family 1, where $N$ is the number of samples in the training set. Despite using optimized software, the former growth is still quite large, and we opt to search only through a fixed number of values of $m$ in families 2--5 that includes $m = 1$, $m=w$, and $m$ close to $w/4$, $w/2$ and $3w/4$ respectively.

\section{Trainability}

Trainability of classifiers is their ability to capture class distributions over training data. Significant failure to classify training data is an indication that the classifier is not flexible enough. Training errors for each class of classifiers is shown in Table \ref{tab:training}.

\begin{table}[!ht]
\begin{center}
\begin{tabular}{|l|c|c|c|c|c|}
\hline
& 1 & 2 & 3 & 4 & 5 \\
\hline
{\tt aa-ao} & 232  & 213  & 223 & 224 & 220 \\
{\tt aa-dcl} & 1 & 1 & 0  & 1 &  1\\
{\tt aa-iy} & 0 &  0 &  0  & 0 &  0\\
{\tt aa-sh} &  1& 0 &  0 & 0 &  0 \\
{\tt ao-dcl} & 1 & 1 & 1 & 1 & 1 \\
{\tt ao-iy} & 0 & 0 &  0 & 0 &  0\\
{\tt ao-sh} & 1 & 0 &  0 & 0 &  0\\
{\tt dcl-iy} & 69  & 47 &  71 & 73& 73 \\
{\tt dcl-sh} & 1 &0  &  0 & 0& 0 \\
{\tt iy-sh} & 8 & 0 &  0 & 0 &  1 \\
\hline
\end{tabular}
\end{center}
\caption{Total number of errors on training data of the best classifiers in the family given in the column for a pair of phonemes given by the row}
\label{tab:training}
\end{table}
From the table we can see that family 1 of classifiers is least trainable, which can be expected, since it is subsumed by family 2. On the other hand, family 2 is slightly more trainable than others.

\section{Performance on test data}

The crucial characteristic of any classifier is its performance on the test data. Table \ref{tab:testing} summarized results of best classifiers within a given family (priority is given to $c(f)$ and $\rho(f)$ is used in case of ties).
\begin{table}[!ht]
\begin{center}
\begin{tabular}{|l|c|c|c|c|c|c|}
\hline
& 1 & 2 & 3 & 4 & 5 & Family 2 pctg.\\
\hline
{\tt aa-ao} & 95  & 91 & 96 & 98 & 94 & 79.95\% \\
{\tt aa-dcl} & 0 &0  & 0 & 0 &  0 & 100\%\\
{\tt aa-iy} & 0 & 0 & 0 & 0 &  0 & 100 \% \\
{\tt aa-sh} & 2 & 1 &0 & 0 &  0 & 99.75\% \\
{\tt ao-dcl} & 0 &0  & 1 & 0 & 0 & 100\% \\
{\tt ao-iy} & 0 & 0 & 1 & 2&  1 & 100 \% \\
{\tt ao-sh} & 1 & 0 & 0 & 0 &  0 & 100 \% \\
{\tt dcl-iy} & 30 & 16 &24 &25 & 34 & 96.84\% \\
{\tt dcl-sh} &1  & 1 & 0&0  & 0 & 99.76\% \\
{\tt iy-sh} &  5& 1 & 2& 2&  1 & 99.81\% \\
\hline
\end{tabular}
\end{center}
\caption{Total number of errors on test data of the best classifiers in the family given in the column for a pair of phonemes given by the row}
\label{tab:testing}
\end{table}
We can again see that family 2 of classifiers provides the best testing performance. Intriguing is poor performance of family 5 on discrimination of {\tt dcl} versus {\tt iy}. In fact, there is a simple classifier in family 5 with only 24 errors on test data. Putting priority on the correct train count $c(f)$ rather than on $\rho(f)$ resulted in reporting performance of poorer classifier ({\tt dcl.iy.2.disc} in R code listing below) rather than the better one ({\tt dcl.iy.1.disc}).

\begin{lstlisting}[language=R]
dcl.iy.1.left.1 = function(x) max(x[2:6])
dcl.iy.1.right.1 = function(x) max(x[10:13])
dcl.iy.1 = function(x) dcl.iy.1.left.1(x) - dcl.iy.1.right.1(x)
dcl.iy.1.disc = function(x) if (dcl.iy.1(x) < 0) "iy" else "dcl"

dcl.iy.2.left.1 = function(x) max(x[1:4])
dcl.iy.2.right.1 = function(x) Q(x,8,14,5)
dcl.iy.2 = function(x) dcl.iy.2.left.1(x) - dcl.iy.2.right.1(x)
dcl.iy.2.disc = function(x) if (dcl.iy.2(x) < 0) "iy" else "dcl"

Q = function(x,a,b,p) { v = sort(x[a:b]); return(v[p])}
\end{lstlisting}

\section{Visualization}

The families of discriminators we have examined in this article can be readily visualized. 

\begin{figure}[!ht]
\includegraphics[width=0.47\textwidth]{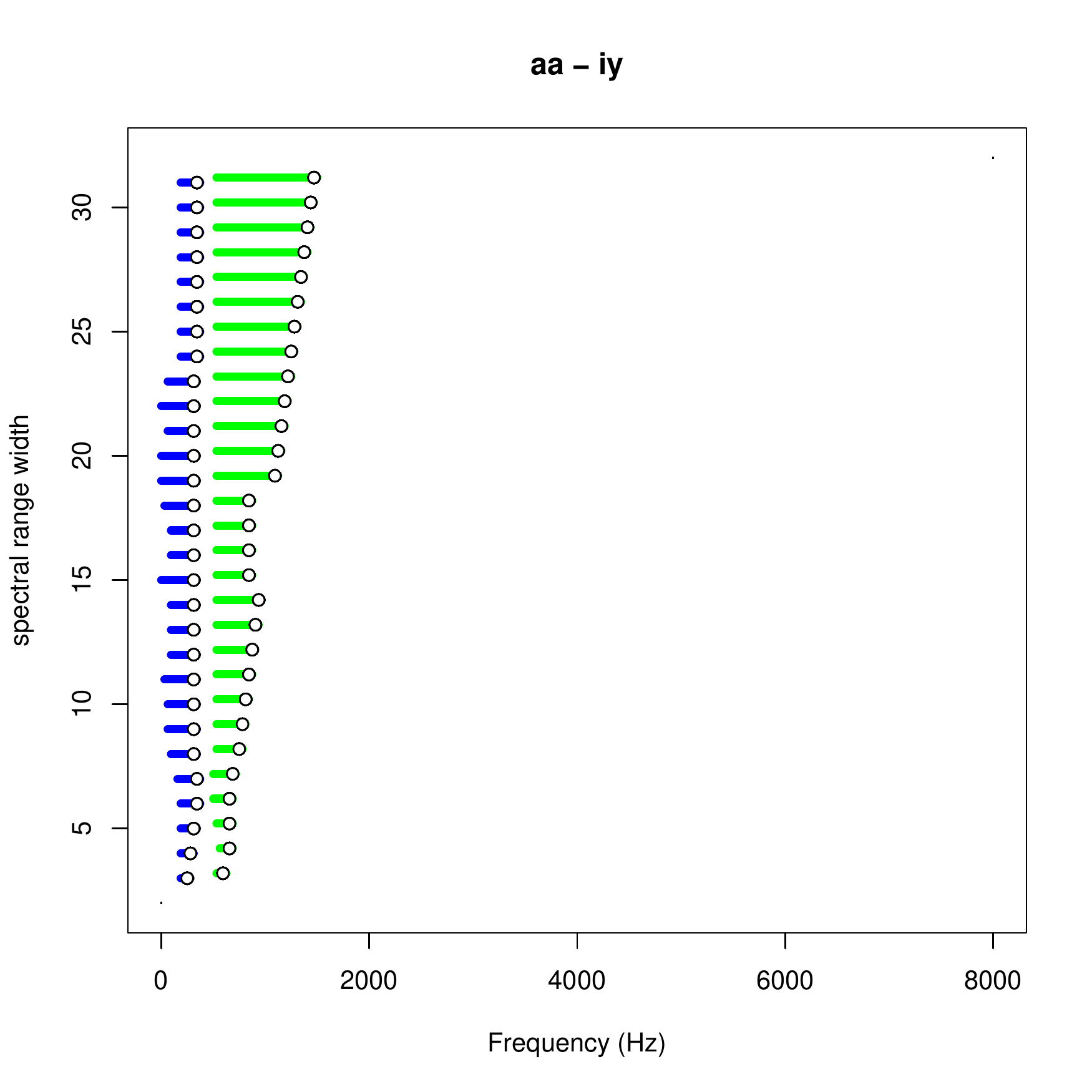}
\includegraphics[width=0.47\textwidth]{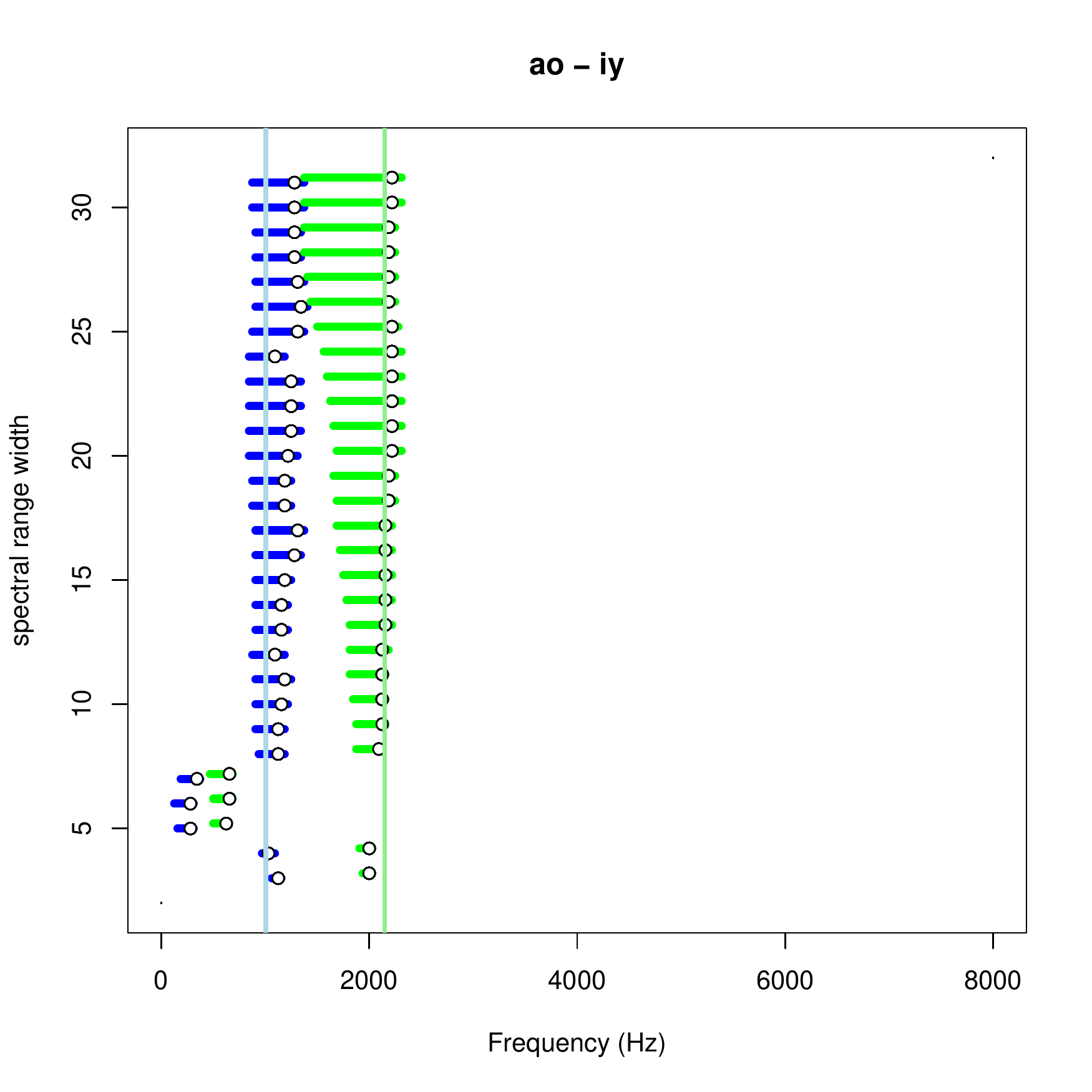}
\caption{Supports and position of parameter $m$ (white circles) of optimal classifiers of a given width in family 5. In the right picture locations of average frequencies for formats F2 for ao (light blue) and iy (light green) are indicated by vertical lines. The average values were taken from \cite{Formanty}.
}
\label{fig:supports}
\end{figure}

In Figure \ref{fig:supports} we can see how support of $f_1$ and $f_2$ changes if we allow increased size of its support. In the right picture we can see that support of components of $f$ matches well with formants.

\section{Conclusion}
We have conducted a search for structure of optimal $Z$-classifiers in 5 families of functions in $KS$-algebra. Among families we considered, slightly better results on both training and test data were obtained in family 2. We have demonstrated that classifiers found by our procedure reflect well known formant concept. Advantages of these classifiers include clear interpretation, visualizations and lack of any continuously varied parameters resulting in low Kolmogoroff's complexity. 

Further research should investigate more general classes of $KS$-algebra based classifiers, namely $B$ and $A$--classifiers, adjustments for psychoacoustic phenomena, and develop means to compose single feature classifiers.
%
%
\bibliographystyle{elsarticle-num}
\bibliography{alg_bib}

\end{document}